\pdfoutput=1

\documentclass[11pt]{article}

\usepackage[final]{coling}

\usepackage{times}
\usepackage{latexsym}

\usepackage[T1]{fontenc}

\usepackage[utf8]{inputenc}
\usepackage{amssymb}
\usepackage{amsmath}

\usepackage{microtype}

\usepackage{inconsolata}

\usepackage{graphicx}
\usepackage{subcaption}
\usepackage{enumitem}

\title{Multi-Graph Co-Training for Capturing User Intent in Session-based Recommendation}
\author{
  \textbf{Zhe Yang\textsuperscript{1}}, 
  \textbf{Tiantian Liang\textsuperscript{1}\thanks{Corresponding author.}} \\
  \textsuperscript{1}School of Computer Science and Technology, Soochow University, Suzhou, China \\
  \texttt{yangzhe@suda.edu.cn, ttliang2023@stu.suda.edu.cn}
}

\begin{document}
\maketitle
\footnotetext[1]{Zhe Yang and Tiantian Liang are joint first authors with equal contributions to this work.}
\begin{abstract}
Session-based recommendation focuses on predicting the next item a user will interact with based on sequences of anonymous user sessions. A significant challenge in this field is data sparsity due to the typically short-term interactions. Most existing methods rely heavily on users' current interactions, overlooking the wealth of auxiliary information available. To address this, we propose a novel model, the Multi-Graph Co-Training model (MGCOT), which leverages not only the current session graph but also similar session graphs and a global item relation graph. This approach allows for a more comprehensive exploration of intrinsic relationships and better captures user intent from multiple views, enabling session representations to complement each other. Additionally, MGCOT employs multi-head attention mechanisms to effectively capture relevant session intent and uses contrastive learning to form accurate and robust session representations. Extensive experiments on three datasets demonstrate that MGCOT significantly enhances the performance of session-based recommendations, particularly on the Diginetica dataset, achieving improvements of up to 2. 00\% in P @ 20 and 10. 70\% in MRR @ 20. Resources have been made publicly
available in our \href{https://github.com/liang-tian-tian/MGCOT}{GitHub} repository \url{https://github.com/liang-tian-tian/MGCOT}.
\end{abstract}

\section{Introduction}
Session-based recommendation aims to discover user intent by learning from the sequence of items in the current session, ultimately recommending items of interest to the user. A session typically refers to a sequence of user interactions with multiple items within a period of time, such as consecutively clicking on several products on a shopping platform. Session-based recommendation is particularly effective in attracting and retaining anonymous users, especially those who prioritize privacy or are first-time users of the platform. This approach is crucial for e-commerce platforms and streaming services, such as Amazon or YouTube. However, the greatest challenge in session-based recommendation is severe data sparsity, as it primarily focuses on the user actions within the current session and fails to adequately capture the intrinsic relationships between items and the similar intents across different sessions. \par
Early session-based recommendation models leverage the Markov chain assumption \cite{rendle2010factorizing} to capture sequential patterns. With advances in neural networks, Recurrent Neural Networks (RNNs) \cite{ hidasi2015session, li2017neural, ren2019repeatnet} are employed to extract item transition relationships using recurrent units or attention layers.
Graph Neural Networks (GNNs) \cite{wu2019session} convert session sequences into graph structures to capture higher-order item relationships. While GNNs outperform in capturing pairwise item transitions, they may have weaker long-term dependencies. Graph Attention Networks (GATs) \cite{wang2019collaborative, wang2020global} address this issue by incorporating attention mechanisms, but their high memory consumption limits their application to current session data, often neglecting global item correlations. Without attention mechanisms, recommendation precision may decline.\par
Self-Supervised Learning (SSL) \cite{liu2021self, xia2021self, xia2021selfco} provides effective solutions for data sparsity by constructing both global and local graphs to enhance session representations. However, these methods often fail to capture similar intents across different sessions, leading to incomplete information modeling. \par
More recently, MiasRec\cite{choi2024multi} generates multiple session representations centered around each item and dynamically selects the most relevant ones to capture user intent. This approach performs well in longer-session contexts, but its effectiveness diminishes in shorter-session scenarios. \par
To address these issues, we propose a multi-graph co-training model with various attention mechanisms that captures user intent from different views and filters out irrelevant items. Our model includes tree views to obtain the session representation: the current view, which reflects item transitions within the current session; the local view, which captures relationships between similar sessions; and the global view, which encompasses item relationships across all sessions. Each view includes an encoding layer, implemented with either Gated Recurrent Unit (GRUs) or GNNs, along with attention mechanisms to generate session embeddings. Finally, contrastive learning is applied between the combination of current and local graphs and the global graph to capture more accurate session representations. \par
In summary, the main contributions of this paper are as follows: \par
\begin{itemize}
\item We introduce the construction of a frequency-based current item graph and employ shortest path algorithms in the global graph to further enhance the model's capacity to comprehensively transform session data into graph representations.
\item We introduce various attention mechanisms to effectively capture session information. These mechanisms allow the model to extract relevant data from diverse aspects of the session and emphasize critical information, thereby aligning more closely with the user's intent. 
\item We propose a co-training approach between the combination of current and local graphs and the global graph using contrastive learning, enabling a more comprehensive and complementary understanding of user intent from different views. 
\item We conduct extensive experiments on three real-world datasets, demonstrating that MGCOT outperforms SOTA models.  Specifically, MGCOT achieves a 5.02\% increase in M@10 on Tmall, a 2.17\% increase on RetailRocket, and a 10.53\% increase on Diginetica.
\end{itemize}

\section{Related Work}
In this section, we introduce the related work of our model MGCOT, which includes GNN-based methods and self-supervised learning.
\subsection{GNN-based Methods}
GNNs \cite{wu2020comprehensive} have been widely used in capturing complex transition relationships and have demonstrated substantial effectiveness. Sessions can be well represented as graphs, and various studies have explored how GNNs can enhance session recommendation accuracy. The SR-GNN model \cite{wu2019session} is the first to utilize the Gated Graph Neural Network (GGNN) for learning item embeddings by propagating information on the session graph. Qiu et al. \cite{qiu2019rethinking} propose the FGNN model, which leverages multi-head attention to aggregate information from an item's neighborhood. GC-SAN \cite{xu2019graph} is an evolution of the SR-GNN, which applies a self-attention mechanism to model item co-occurrences. GCE-GNN \cite{wang2020global} aggregates item information from both item-level and session-level through graph convolution and self-attention mechanism. MGIR \cite{han2022multi} models not only sequential and global co-occurrence relations but also incompatible relations within a graph. KMVG\cite{chen2023knowledge} utilizes multi-view graph neural networks and a knowledge graph to more accurately capture user intent. MSGAT \cite{qiao2023bi} introduces a bi-channel model with multiple sparse graph attention networks that takes into account the effects of session intent and noise items. In the GNN-based session recommendation models, multi-graph models have shown significant advantages over single-graph models. This has inspired using a multi-graph co-training model with attention mechanisms to capture session intent more comprehensively. 
\subsection{Self-supervised Learning}
In recent years, SSL has proven to be effective for recommendation. $S^3$-Rec \cite{zhou2020s3} uses the mutual information maximization principle to learn the underlying relationships among items, attributes, and sequences. $S^2$-DHCN \cite{xia2021self} employs a contrastive learning mechanism to enhance hyper-graph modelling through a different line of GCN models. COTREC \cite{xia2021selfco} proposes constructing session data into two views to capture the internal and external connectivity of sessions. CGL \cite{pan2022collaborative} integrates SSL with supervised learning to explore correlations across different sessions, thereby improving item representations. HGCMA\cite{chen2023heterogeneous} employs random masking and contrastive learning to learn discriminative node representations in heterogeneous graphs. While SSL has demonstrated great performance in capturing user intent from various views, these methods overlook the potential benefits of combining similar session intents to achieve a more comprehensive and accurate understanding.
\section{Methods}
\begin{figure*}[t]
  \includegraphics[width=0.96\linewidth]{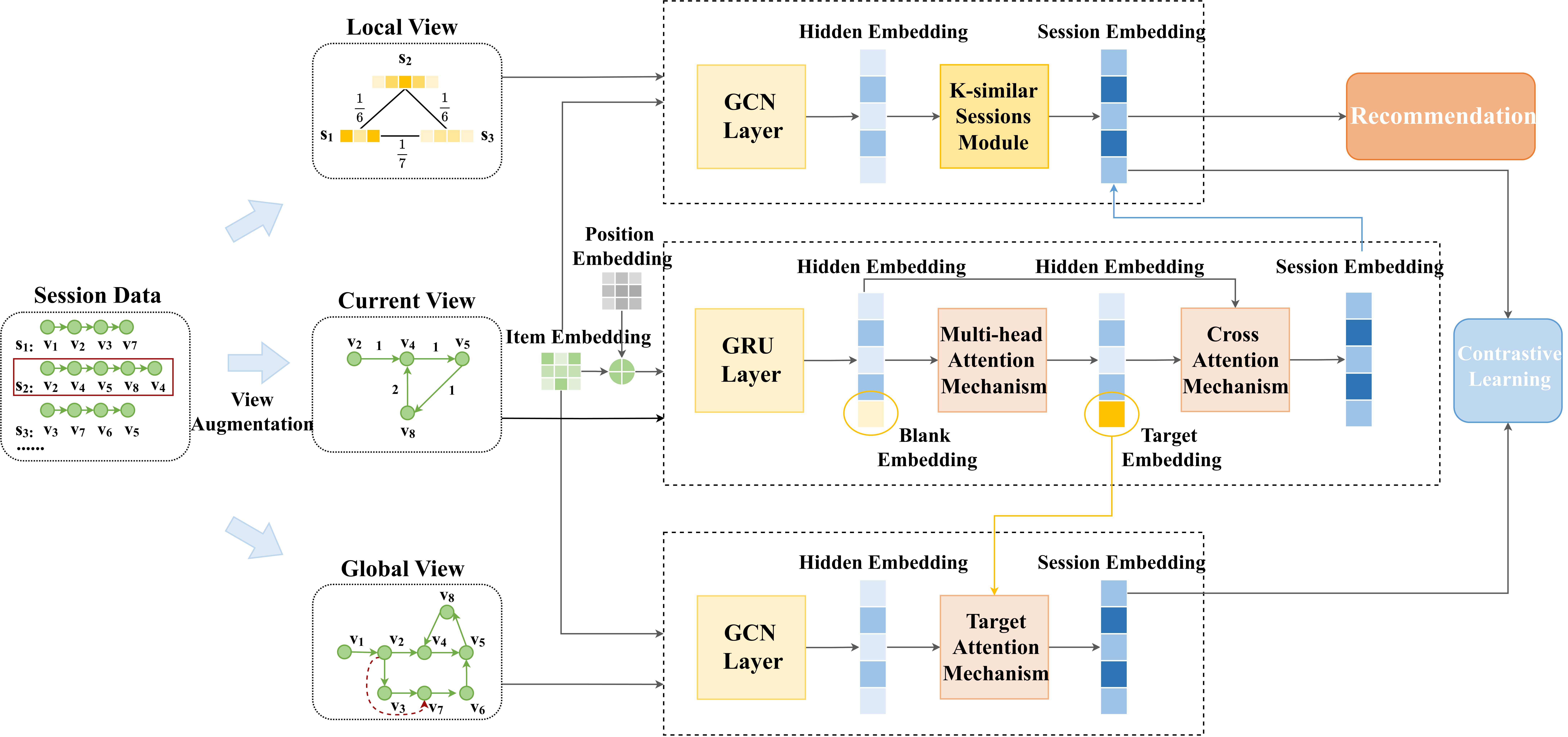} 
  \caption{An overview of the proposed MGCOT framework.}
  \label{fig:overallModel}
\end{figure*}
The model learns session representations from three views: current view, local view, and global view, as illustrated in Figure~\ref{fig:overallModel}. In the current view, we adopt the SR-GNN approach\cite{wu2019session}, utilizing the Gated Graph Neural Network (GGNN) as the initial encoder. This method combines the strengths of Graph Neural Networks (GNNs) and Gated Recurrent Units (GRUs), enabling the model to effectively capture item relationships within a session and extract the session's intent. In the local view, the model first generates session representations for the current batch. It then refines the session representation from the current view by integrating it with representations from the top-k similar sessions identified within the local context. The enhanced session representation is subsequently utilized for the primary recommendation task. In the global view, the model generates the current session representation by extracting item information from all sessions. We also incorporate various attention mechanisms to extract crucial information in these views. \par
To further improve the model's ability to capture item relationships, a contrastive learning approach is employed. This approach compares the enhanced session representation, which integrates current session representation with similar session information from the local view, with the session representation generated from the global view. The method of enhancing the current session representation by integrating similar session information in the local view is inspired by \cite{qiao2023bi}. \par
In this section, we focus on introducing our key ideas: graph construction from different views, various attention mechanisms, and contrastive learning. \par
\subsection{Problem Definition}
We represent the set of sessions in the dataset as $S=\{s_{1},s_{2},\dots,s_{M}\}$, where $M$ denotes the total number of sessions. The set of all items is defined as $V=\{v_{1},v_{2},\dots,v_{N}\}$, where $N$ is the total number of items in the dataset. Each session $s_{t}$ is generated by an anonymous user interacting with a set of items. The session at time $t$ is denoted as $s_{t}=\{v_{1},v_{2},\dots,v_{L}\}$, where $L$ is the length of the current session. The objective of session-based recommendation is to capture the user intent based on the first $L$ consecutive interactions and predict the $L$+1-th potential interaction item.
\subsection{Graph Construction from Different Views}
To fully leverage the available information, we explore relationships between items and sessions from three views: the current view, global view, and local view. These views focus on item relationships within the current session, across all sessions, and among batch sessions. To capture the intrinsic correlations, we first convert sessions into graphs. We propose the current frequency item graph, the global shortest-path item graph, and introduce the local session graph.
\subsubsection{Current Frequency Item Graph}
In session-based recommendation, the order in which users click on items reflects changes in their interests. Sequence information helps the model better understand the current context of the user, thereby improving recommendation accuracy. However, in traditional directed graph construction, different sessions composed of the same items might generate identical graph structures. For example, session sequences $s_1=\{v_2,v_4,v_5,v_5,v_4,v_4\}$ and $s_2=\{v_2,v_4,v_4,v_5,v_5,v_4\}$ may result in indistinguishable graphs. This can lead to a loss of critical sequential information, negatively impacting recommendation results. \par
\begin{figure}[t]
 \centering
  \includegraphics[width=0.48\columnwidth]{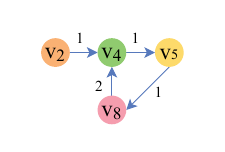}
  \caption{Session Graph with Frequency Indegree}
  \label{fig:session_graph}
\end{figure}

To preserve as much information from the original sessions as possible, we propose a method for constructing directed graphs based on the frequency of item occurrences within the current session. In this method, the in-degree frequency of an item is used as the edge weight in the directed graph structure. For instance, in the session sequence $s_3=\{ v_2,v_4,v_5,v_8,v_4 \}$, the edge from $v_2$ to $v_4$ has a weight of 1, while the edge from $v_8$ to $v_4$ has a weight of 2, as shown in Figure \ref{fig:session_graph}. By introducing frequency-based weights, this method effectively reduces information loss during the session graph construction process, ensuring that sessions like $s_1$ and $s_2$, which differ in edge weights, generate distinct graph structures. This approach better preserves and allows the model to learn the comprehensive information within sessions.

\begin{figure}[t]
 \centering
 \begin{subfigure}[b]{0.48\columnwidth}
  \includegraphics[width=\linewidth]{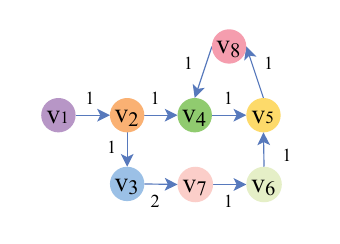}
  \caption{Occurrence Weight Graph}
  \label{fig:global_weight}
  \end{subfigure}
  \hfill
  \begin{subfigure}[b]{0.48\columnwidth}
  \includegraphics[width=\linewidth]{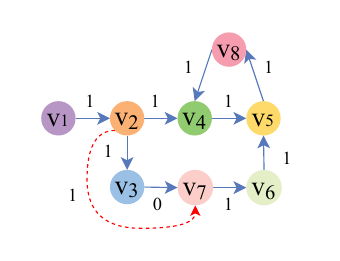}
  \caption{Shortest-path Graph}
  \label{fig:global_shortest}
  \end{subfigure}
  \caption{Global Shortest-path Item Graph}
\end{figure}
\subsubsection{Global Shortest-path Item Graph}
Most existing models based on GNNs perform poorly in capturing long-range dependencies because GNNs only aggregate information from neighboring nodes in each layer. By stacking multiple layers of GNNs, the model can gradually aggregate information from more distant neighbors, but too many layers may lead to overfitting. \par
To address the issue of capturing relationships between distant nodes, we construct the global item graph (as shown in Figure \ref{fig:global_weight}) and then use Dijkstra's algorithm to compute the shortest path between pairs of nodes (as shown in Figure \ref{fig:global_shortest}). First, we transform the weight of each edge to its inverse weight by subtracting the edge weight from the maximum weight of all edges to obtain the corresponding cost value $c_{ij}$. Then, for each node in the global item graph, we calculate the shortest path from that node to all other nodes based on the minimum total cost of all edges along the path. Finally, the calculated cost values are inverted back into weights, allowing the global graph based on the shortest paths to effectively capture the relationships between distant nodes.
In the global graph, the minimum cost matrix $\hat{C}$ and the final edge weight $\hat{w}_{ij}$ are defined as follows:
\begin{equation}
  \label{eq:dijkstra}
  d_{ij} = min(d_{ij}, d_{ik}+c_{kj})
\end{equation}
\begin{equation}
  \label{eq:globalWeight}
  w_{ij} = max(\hat{C}+1)-\hat{c}_{ij}
\end{equation}
Here, $d_{ij}$ represents the current shortest distance from the start node $i$ to node $j$, and $d_{ik}+c_{kj}$ represents the total cost of the path from the start node $i$ to node $j$ via node $k$. The Equation~\ref{eq:dijkstra} indicates that if the cost of reaching node $j$ through node $k$ is less than the currently known shortest path, the shortest path value is updated accordingly. In Equation~\ref{eq:globalWeight}, $\hat{C}$  represents the matrix of minimum costs for all edges, and $\hat{c}_{ij}$ denotes the minimum cost from node $i$ to node $j$.
\subsubsection{Local Session Graph}
When constructing a local session graph, we follow the method described in \cite{qiao2023bi}. Each session is treated as a node in the graph. The edges between nodes are determined by calculating the Jaccard similarity of the items shared between sessions. A higher Jaccard similarity indicates a greater overlap of items between the two sessions, resulting in a higher edge weight. This indicates the intent of the two sessions is more similar.

\subsection{Attention Mechanisms}
The attention mechanism can effectively capture important information related to the intent of the session. In this paper, we focus on two main types of attention mechanisms: the multi-head attention mechanism and the target attention mechanism. The multi-head attention mechanism is used to comprehensively capture significant information within the current session, while the target attention mechanism extracts information from a global view and learns information related to the target item by incorporating the context from the current view. Additionally, we incorporate the cross attention mechanism as described in \cite{qiao2023bi}.
\subsubsection{Multi-head Attention Mechanism}
The multi-head attention mechanism uses multiple independent attention heads to compute attention scores in different subspaces simultaneously. This approach allows the model to focus on various aspects of the input sequence within the same layer. The self-attention mechanism is inherently global, enabling each position's output vector to interact with every other position in the input sequence, effectively capturing long-range dependencies. In the current view, the multi-head self-attention mechanism effectively captures relevant information in the session. \par
The multi-head self-attention mechanism mainly consists of three parts. First, a feedforward neural network is employed to enhance the representation of the query vector $Q$, making it more flexible and general, thus distinguishing it from the key vector $K$ and the value vector $V$. Here, $H_t$ denotes the initial embedding of the session in the current view.
\begin{equation}
  \label{eq:multi_Q}
  \hat{Q} = f(H_tW_Q+b_Q) 
\end{equation}
where $W_Q \in \mathbb{R}^{2d \times 2d}$ is the weight matrix, $b_Q \in \mathbb{R}^{2d}$ is the bias vector, and $f(\cdot)$ denotes the ReLU activation function. \par
Second, a sparse transformation is applied to generate attention weights, ensuring that the new item embeddings are more relevant to the target item embeddings. The attention weights are calculated as follows:
\begin{equation}
  \label{eq:multi_alpha_t}
\alpha_t = \sigma(W_{\alpha_t}h_t+b_{\alpha_t})+1
\end{equation}
where $W_{\alpha_t} \in \mathbb{R}^{1 \times d_k}$ is the weight matrix, $b_{\alpha_t} \in \mathbb{R}^{d_k}$ is the bias vector, $d_k$ is the dimension for each attention head, and $\sigma$ denotes the sigmoid activation function. The vector $h_t$ is a special item index added to the end of the input sequence to indicate the item to be predicted. This special item embedding helps the model capture the overall session pattern rather than focusing solely on individual item characteristics.\par
Finally, the score of the multi-head attention mechanism is computed as follows:
\begin{equation}
  \label{eq:multi_HmultiAtt_k}
H_{cur}^k = \alpha_{t} \mbox{-} entmax(\frac{\hat{Q}K^T}{\sqrt{d_k}})V 
\end{equation}
\begin{equation}
  \label{eq:multi_HmultiAtt}
H_{cur} = Concat(H_{cur}^1, H_{cur}^2,..., H_{cur}^{H_n})
\end{equation}
where $\hat{Q}$ is the mapped representation of the current session, $K$ and $V$ are the key and value representations of the current session, $d_k$ is the dimension of each attention head, $H_{cur}^k$ is the output of the $k$-th attention head, and entmax is a sparse attention mechanism. $H_n$ represents the number of attention heads. \par
Although the multi-head attention mechanism learns new representations for all items, it is primarily based on linear projections. Subsequently, a feedforward neural network is applied to learn more nonlinear features:
\begin{equation}
\resizebox{0.4\textwidth}{!}{$\hat{H}_{cur}=Dropout(W_2(f(W_1 H_{cur} + b_1)) + b_2) + H_{cur}$}
\label{eq:multi_Hs}
\end{equation}
where $W_1, W_2 \in \mathbb{R}^{2d \times 2d}$ are weight matrices, $b_1, b_2 \in \mathbb{R}^{2d}$ are bias vectors, and $f(\cdot)$ represents the ReLU activation function. $\hat{H}_{cur}=\{h_{1^{\prime}}, h_{2^{\prime}}, ..., h_{t^{\prime}}\}$ represents the output processed by the multi-head self-attention mechanism, where $h_{t^{\prime}}$ is the learned target item embedding. The dropout layer is included to prevent overfitting, while residual connections and layer normalization are applied to mitigate instability during training.
\subsubsection{Target Attention Mechanism}
The target attention mechanism aims to learn the representation of the entire session based on the learned target embeddings and initial inputs. It adjusts weights to reduce noise in the current initial session representation $H_g$ from the global view.
First, the target attention weights are computed as follows:
\begin{equation}
  \label{eq:target_alpha}
\alpha_s = \sigma(W_{\alpha_s}h_{t^{\prime}}+b_{\alpha_s})+1 
\end{equation}
where $W_{\alpha_s} \in \mathbb{R}^{1 \times 2d}$ is the weight matrix, $b_{\alpha_s} \in \mathbb{R}^{2d}$ is the bias vector, $\sigma$ denotes the sigmoid activation function, and $h_{t^{\prime}}$ is the representation of the target item obtained through the multi-head attention mechanism in the current view.
The attention weight $w_s$ is calculated as follows:
\begin{equation}
  \label{eq:target_ws}
\resizebox{0.4\textwidth}{!}{$ w_s=\alpha_s \mbox{-}entmax(W_0f(W_1H_{g}+W_2h_{t^{\prime}}+b_{0}))$} 
\end{equation}
where $W_1, W_2 \in \mathbb{R}^{2d \times 2d}$ are weight matrices, $W_0 \in \mathbb{R}^{1 \times 2d}$ is a weight matrix, $b_{0} \in \mathbb{R}^{2d}$ is a bias vector, $f(\cdot)$ is the ReLU activation function.
Finally, the final session representation $\hat{H}_g$ in the global view is computed as:
\begin{equation}
  \label{eq:target_Hg}
\hat{H}_g = \sum_{k=0}^{n}w_sh_g^k
\end{equation}
where $h_g^k \in H_g$ denotes the representation of each item in initial session embeddings.

\subsection{Contrastive Learning}
The core idea of contrastive learning is to build better feature representations by learning the similarities between similar samples and the differences between dissimilar samples. Specifically, for each session, we use the session representation from the current and local views, denoted as $\hat{H}_r^b$, which integrates information from other similar sessions within the batch. This representation is then contrasted with the session representation obtained from the global view, denoted as $\hat{H}_g^b$. During training, we treat the representations of the same session from different views within the same batch as positive samples, aiming to pull these positive samples closer together. Conversely, we treat the representations of other sessions within the same batch as negative samples, aiming to push them further away from the current session representation.
The similarity scores for positive and negative samples are calculated as follows:
\begin{align}
\label{eq:Contrastive_posNeg}
\text{Sim}_p &= \hat{H}_r^b \cdot \hat{H}_g^b \\
\text{Sim}_n &= \hat{H}_r^b \cdot \hat{H}_{g_{\text{shuffled}}}^b
\end{align}
where $\hat{H}_{g_{\text{shuffled}}}^b$ represents the global view representations of other sessions in the batch, excluding the current session and randomly shuffled.
The contrastive learning loss is computed as:
\begin{equation}
  \label{eq:Contrastive_loss}
\resizebox{0.4\textwidth}{!}{$L_{\text{contrastive}} = -log(\sigma(\frac{\text{Sim}_p}{\tau}))-log(\sigma(-\frac{\text{Sim}_n}{\tau}))$}
\end{equation}
where $\tau$ is a temperature parameter used to scale the similarity scores to enhance the effectiveness of contrastive learning. The main recommendation encoder uses the cross-entropy loss function, defined as:
\begin{equation}
  \label{eq:Main_loss}
\resizebox{0.4\textwidth}{!}{$L_{\text{main}} =-\sum_{i=1}^Ny_ilog(\hat{y}_i)+(1-y_i)log(1-\hat{y}_i)$}
\end{equation}
where $\hat{y}_i$ denotes the probability of item $v_i$ being the next click in the current session, and $y_i$ is a binary label that equals 1 if item $v_i$ is the ground truth next click and 0 otherwise. 
The total loss function is defined as:
\begin{equation}
  \label{eq:Total_loss}
L = L_{\text{main}} + \beta L_{\text{contrastive}}
\end{equation}
where $\beta$ is a hyperparameter used to control the extent of contrastive learning.

\section{Experiments}
\subsection{Datasets}
We evaluate our model using three real-world benchmark datasets: Tmall, RetailRocket and Diginetica. The details of these datasets are presented in Table \ref{tab:datasets}. The Tmall dataset contains user shopping logs and is provided by the IJCAI-15 competition. The RetailRocket dataset, released by an e-commerce company on Kaggle, includes six months of user browsing activities. The Diginetica dataset consists of typical transaction data from the CIKM Cup 2016. \par
To ensure data quality and relevance, we preprocess the data as follows\cite{wu2019session, wang2020global, xia2021selfco}: We exclude sessions with a length of 1 and remove items that appear fewer than 5 times. Datasets are split with the most recent data as the test set and the rest as the training set. We also enhance the data by segmenting each session and generating labels, where each sequence is paired with the next item as the label. This augmentation improves the model’s ability to learn sequential patterns.

\begin{table}
  \centering
   \caption{Dataset Statistics}
  \resizebox{0.98\columnwidth}{!}{%
    \begin{tabular}{lcccc}
      \hline
      \textbf{Dataset} & 
      \textbf{Train} &
      \textbf{Test} &
      \textbf{Items} &
      \textbf{Avg.Len.} \\
      \hline
      Tmall & 351,268 & 25,898 & 40,728 & 6.69 \\
      RetailRocket & 433,643 & 15,132 & 36,968 & 5.43 \\
      Diginetica & 719,470 & 60,858 & 43,097 & 5.12 \\
      \hline
    \end{tabular}
  }
  \label{tab:datasets}
\end{table}

\subsection{Baselines}
To ensure a fair comparison, we select representative models from various categories, including traditional methods such as FPMC \cite{rendle2010factorizing}, RNN-based  models like GRU4Rec \cite{hidasi2015session} and NARM \cite{li2017neural}, GNN-based models such as SR-GNN \cite{wu2019session}, GCE-GNN \cite{wang2020global}, HICN\cite{sun2024exploiting}, Mssen\cite{zheng2024hypergraph} and MGIR \cite{han2022multi}, attention-based models like STAMP \cite{liu2018stamp}, MTAW \cite{ouyang2023mining} and MSGAT \cite{qiao2023bi}, and contrastive learning methods such as $S^2$-DHCN \cite{xia2021self}, to compare with the MGCOT model.

\subsection{Experiment Setting}
Following previous work \cite{qiao2023bi}, we set the batch size to 512, the embedding size to 100, and the $L_2$ regularization to $10^{-5}$. We use the Adam optimizer with a learning rate of 0.001, which decays by a factor of 0.1 every three epochs. Our model uses a single layer of GCN. The number of similar sessions selected is 6 for Tmall, 3 for Diginetica, and 2 for RetailRocket. The scale factor $\beta$ for the contrastive learning loss is set to 0.05 for Tmall, and 5 for both Diginetica and RetailRocket. The number of attention heads $H_n$ in the multi-head attention mechanism is set to 1 for Diginetica, and 2 for Tmall and RetailRocket.

\begin{table*}
  \centering
  \caption{Performances of all comparison methods on three datasets}
  \resizebox{\textwidth}{!}{
  \begin{tabular}{lcccccccccccc}
    \hline
    \textbf{Dataset} & 
    \multicolumn{4}{c}{\textbf{Tmall}} & 
    \multicolumn{4}{c}{\textbf{RetailRocket}} & 
    \multicolumn{4}{c}{\textbf{Diginetica}} \\
    \hline
    \textbf{Method} & 
    \textbf{P@10} & 
    \textbf{M@10} &
    \textbf{P@20} & 
    \textbf{M@20} &
    \textbf{P@10} & 
    \textbf{M@10} &
    \textbf{P@20} & 
    \textbf{M@20} &
    \textbf{P@10} & 
    \textbf{M@10} &
    \textbf{P@20} & 
    \textbf{M@20} %
    \\
    \hline
    \textbf{FPMC(WWW'10)} & 
    13.10 & 7.12 & 16.06 & 7.32 &
    25.99 & 13.38 & 32.37 & 13.82 &
    15.43 & 6.20 & 26.53 & 6.95\\
    \textbf{GRU4Rec(ICLR'16)} & 9.47 & 5.78 & 10.93 & 5.89 & 38.35 & 23.27 & 44.01 & 23.67 & 17.93  & 7.33 & 29.45 & 8.33 \\ 
  
    \textbf{NARM(CIKM’17)} & 19.17 & 10.42 & 23.30 & 10.70 & 42.07 & 24.88 & 50.22 & 24.59 & 35.44 & 15.13 & 49.70 & 16.17 \\

\textbf{STAMP(SIGKDD’18)} & 22.63 & 13.12 & 26.47 & 13.36 & 42.95 & 24.61 & 50.96 & 25.17 & 33.98 & 14.26 & 45.64 & 14.32 \\

\textbf{SR-GNN(AAAI’19)} & 23.41 & 13.45 & 27.57 & 13.72 & 43.21 & 26.07 & 50.32 & 26.57 & 36.86 & 15.52 & 50.73 & 17.59 \\

\textbf{GCE-GNN(SIGIR’20)} & 28.01 & 15.08 & 33.42 & 15.42 & 48.22 & 28.36 & 55.78 & 28.72 & 41.16 & 18.15 & 54.22 & 19.04 \\

\textbf{S2-DHCN(AAAI’21)} & 26.22 & 14.60 & 31.42 & 15.05 & 46.15 & 26.85 & 53.66 & 27.30 & 39.87 & 17.53 & 53.18 & 18.44 \\

\textbf{MGIR(SIGIR’22)} & 30.71 & 17.03 & 36.31 & 17.42 & 47.90 & 28.68 & 55.35 & 29.20 & 40.63 & 17.86 & 53.73 & 18.77 \\

\textbf{MTAW(SIGIR’23)} & 31.67 & 18.90 & 37.17 & 19.14 & 48.41 & 29.96 & 56.39 & 30.52 & - & - & - & - \\

\textbf{MSGAT(CIKM’23)} & \underline{39.21} & \underline{20.92} & \underline{45.43} & \underline{21.35} & \underline{57.00} & \underline{32.73} & \underline{63.68} & \underline{33.21} & \underline{57.09} & \underline{26.30} & \underline{66.97} & \underline{26.91} \\

\textbf{HICN(SDM’24)} & 31.31 & 18.90 & 35.48 & 19.17 & 49.74 & 29.81 & 57.85 & 30.37 & - & - & - & - \\

\textbf{Mssen(LREC-COLING’24)} & 33.53 & 18.98 & 38.51 & 19.60 & - & - & - & - & 42.33 & 19.88 & 55.17 & 19.64 \\

\hline

\textbf{MGCOT} & \textbf{41.28} & \textbf{21.97} & 
\textbf{47.80} & \textbf{22.40} & 
\textbf{57.57} & 
\textbf{33.44} & \textbf{63.78} & \textbf{33.89} & \textbf{58.04} & \textbf{29.07} & \textbf{68.31} & \textbf{29.79} \\

\textbf{Improv.(\%)} & 5.28 & 5.02 & 5.22 & 4.92 & 1.00 & 2.17 & 0.16 & 2.05 & 1.66 & 10.53 & 2.00 & 10.70 \\
\hline
  \end{tabular}
  }
  \label{tab:experiments}
\end{table*}

\subsection{Experiment Results}
Table \ref{tab:experiments} presents the experimental results of the MGCOT model compared to baseline models across three datasets. The best results are highlighted in bold, and the second-best results are underlined.

Experimental results show that traditional machine learning methods like FPMC underperform compared to deep learning approaches. FPMC, which combines matrix factorization and Markov chains, fails to capture long-term dependencies. Among RNN-based models, NARM outperforms GRU4Rec by using attention mechanisms to identify key relationships within sessions. STAMP relies solely on self-attention focused on the last item, replacing RNN encoders with attention layers to better capture short-term interests. MTAW, which models user interests dynamically with an attention mechanism and an adaptive weight loss function, enhances recommendation personalization. Overall, these models demonstrate the effectiveness of attention mechanisms in session-based recommendations.

Recently, GNN-based models have surpassed RNNs by uncovering spatial relationships between items. SR-GNN employs gated GNNs and a self-attention mechanism to capture session embeddings, while GCE-GNN constructs global and local graphs for cross-session learning. $S^2$-DHCN converts sessions into hypergraphs and line graphs using self-supervised learning, inspiring the application of contrastive learning in multi-graph models. MGIR improves session representations by modeling global item relationships, including negative, co-occurrence, and sequential links. HICN boosts performance by leveraging sequential hyperedges and inter-hyperedge modules. Mssen uses multi-collaborative self-supervised learning in hypergraph neural networks to capture high-order relationships and address data sparsity. MSGAT, with dual-channel GNNs and attention mechanisms, excels at modeling both intra- and inter-session information, highlighting the advantages of GNNs with integrated attention.

Compared to the best baseline models, our MGCOT model shows significant performance improvements. By leveraging graph neural networks, attention mechanisms, and contrastive learning, MGCOT effectively captures latent relationships between sessions and items from current, local, and global views.
\begin{table}
  \centering
   \caption{Ablation study of components in MGCOT.}
  \resizebox{0.98\columnwidth}{!}{%
    \begin{tabular}{lcccccc}
    \hline
     Dataset & 
    \multicolumn{2}{c}{Tmall} & 
    \multicolumn{2}{c}{RetailRocket} & 
    \multicolumn{2}{c}{Diginetica} \\
    \hline
      Method & 
      P@20 &
      M@20 &
      P@20 &
      M@20 &
      P@20 &
      M@20 \\
      \hline
      -NeighborSessions & 28.69 & 14.73 & 54.06 & 29.63 & 52.07 & 18.50 \\
      -MultiAttention & 46.85 & 21.52 & 62.89 & 33.25 & 67.87 & 29.47 \\
      -ContrastiveLearning & 47.60 & 21.87 & 63.17 & 33.81 & 67.84 & 29.31 \\
      MGCOT & \textbf{47.80} & \textbf{22.40} & \textbf{63.78} & \textbf{33.89} & \textbf{68.31} & \textbf{29.79} \\
      \hline
    \end{tabular}
  }
  \label{tab:ablation}
\end{table}
\subsection{Ablation Experiments}
To investigate the contribution of each module in MGCOT, we conduct ablation experiments with the following variants: (1) -NeighborSessions, where the fusion of similar session information from the local view is removed; (2) -MultiAttention, where the multi-head attention mechanism on session embeddings from the current view is removed; and (3) -ContrastiveLearning, where contrastive learning between the session embedding generated from the global view and the main session embedding fused from the local view and current view is removed. \par
As shown in Table \ref{tab:ablation}, removing the fusion of similar session information led to a significant drop in evaluation metrics, indicating that similar sessions are as important as similar items in capturing user intent. Furthermore, both the multi-head attention mechanism and contrastive learning improve model performance, demonstrating the importance of assigning different weights to items when capturing session intent and the benefit of understanding session intent from multiple views.

\subsection{Hyperparameter Experiments}
In this hyperparameter experiment, we analyze the sensitivity of the MGCOT model to different parameter settings across datasets. \par

\begin{figure}[t]
 \setlength{\abovecaptionskip}{-10pt} 
 \centering
 \begin{subfigure}[b]{0.48\columnwidth}
  \includegraphics[width=\linewidth]{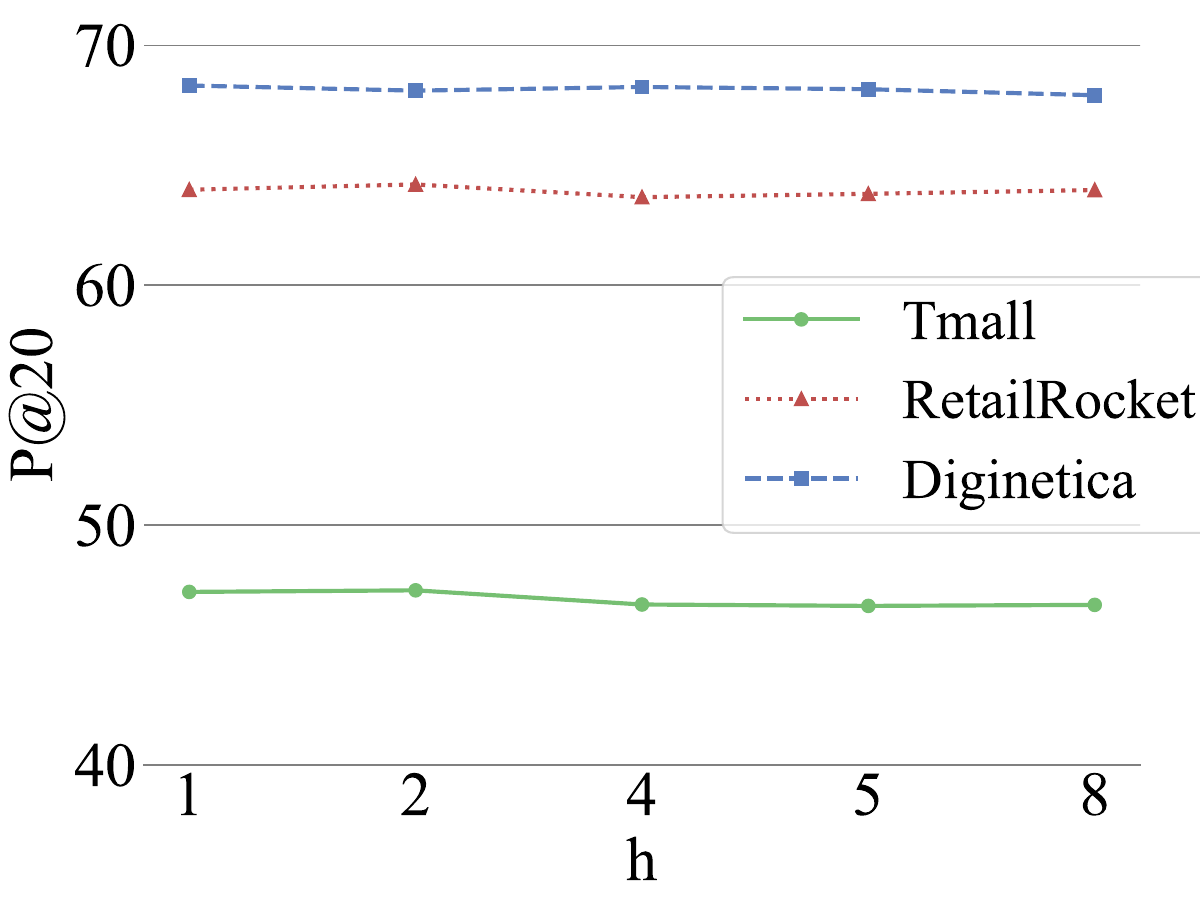}
  \label{fig:P_MultiAttHeads}
  \end{subfigure}
  \hfill
  \begin{subfigure}[b]{0.48\columnwidth}
  \includegraphics[width=\linewidth]{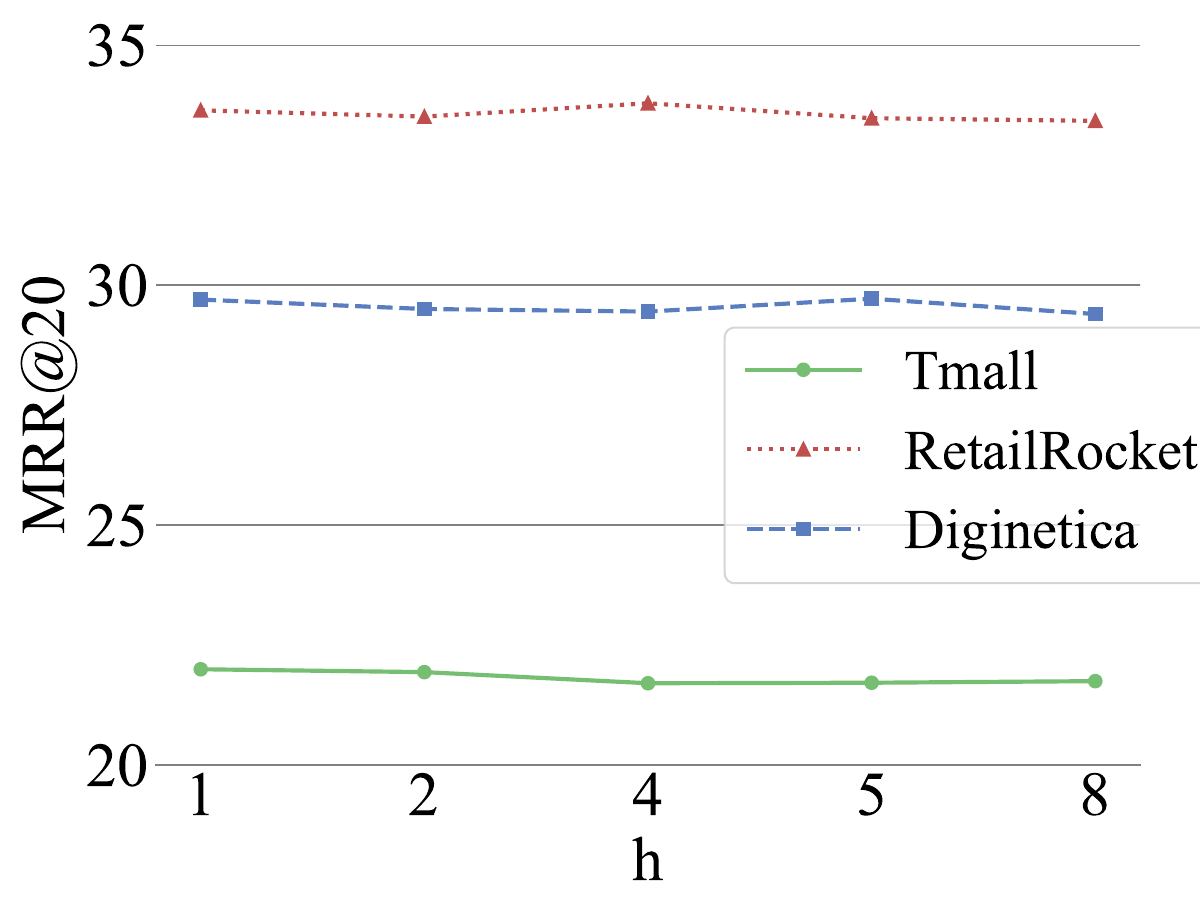}
  \label{fig:MRR_MultiAttHeads}
  \end{subfigure}
  \caption{The number of attention heads $h$}
  \label{fig:MultiAttHeads}
\end{figure}

Figure \ref{fig:MultiAttHeads} shows that we select 2 attention heads for Tmall and RetailRocket, and 1 for Diginetica. In longer sessions, such as those in Tmall, users may experience interest drift, with multiple preference shifts emerging throughout the session. Multiple attention heads are more effective in capturing these varying interests. In shorter sessions, like those in Diginetica, a single attention head is sufficient to capture the main behavioral pattern. \par

\begin{figure}[t]
 \setlength{\abovecaptionskip}{-10pt} 
 \centering
 \begin{subfigure}[b]{0.48\columnwidth}
  \includegraphics[width=\linewidth]{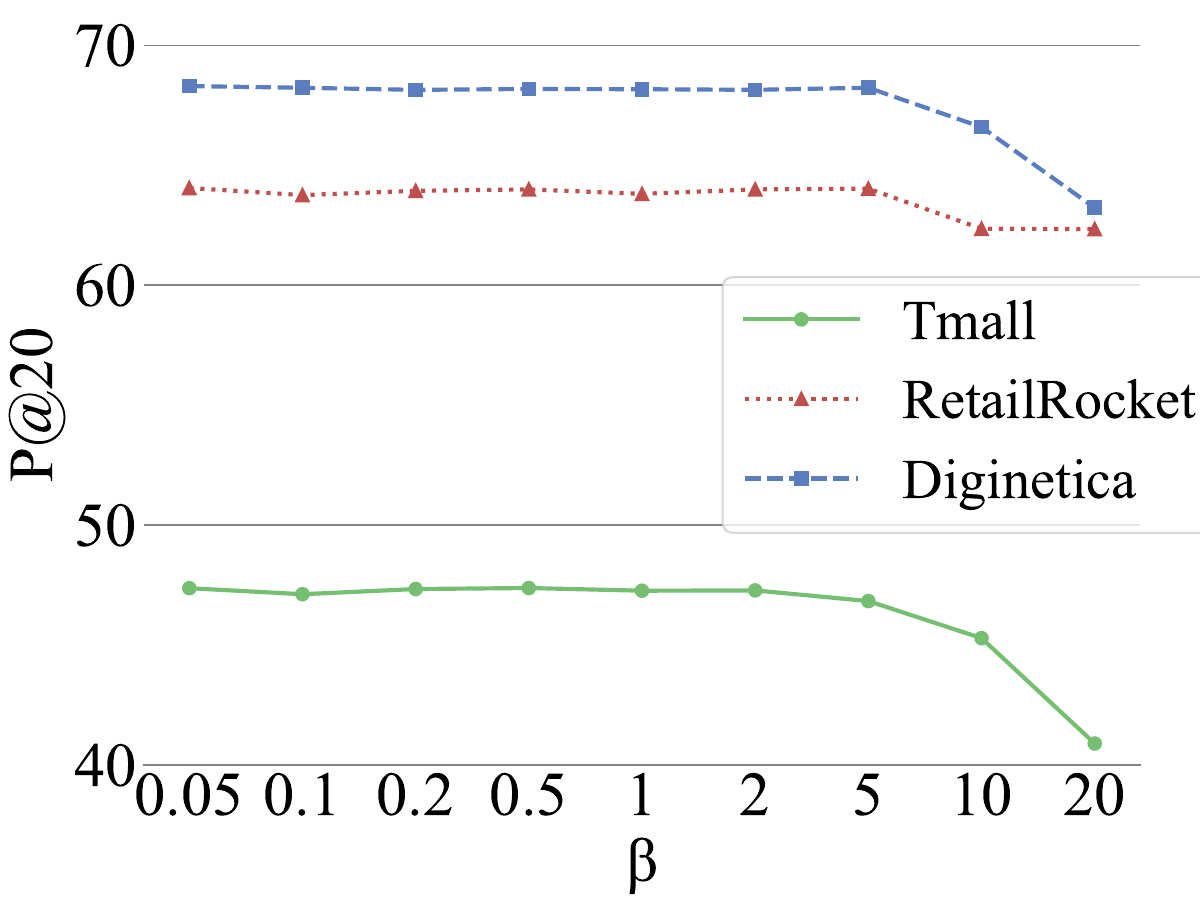}
  \label{fig:P_contrastiveWeight}
  \end{subfigure}
  \hfill
  \begin{subfigure}[b]{0.48\columnwidth}
  \includegraphics[width=\linewidth]{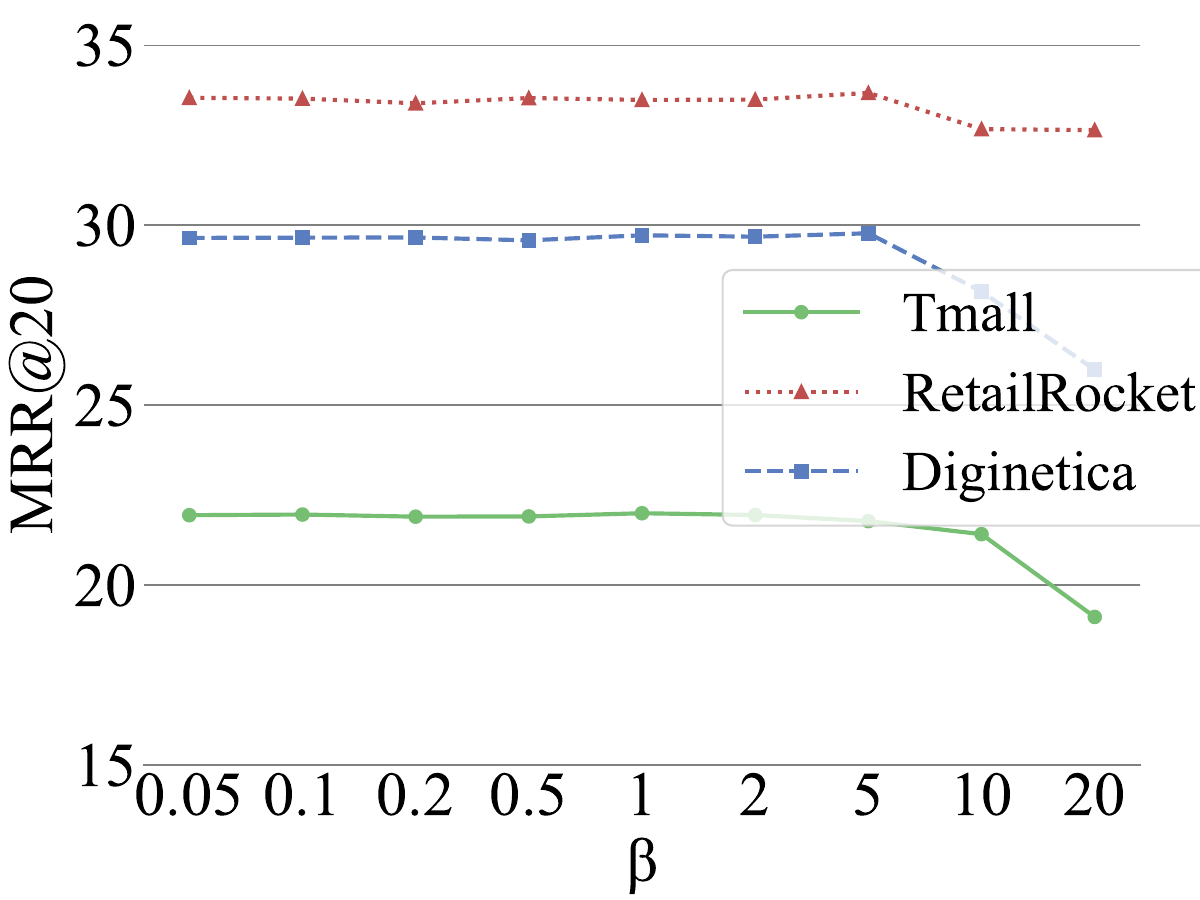}
  \label{fig:MRR_contrastiveWeight}
  \end{subfigure}
  \caption{The weight of contrastive loss $\beta$}
  \label{fig:contrastiveWeight}
\end{figure}

In Figure \ref{fig:contrastiveWeight}, the contrastive loss weight is set to 0.05 for Tmall and 5 for RetailRocket and Diginetica. In the longer sessions of Tmall, users may frequently compare or select similar items, attention mechanisms and similar session fusion more effectively capture user intent, making a lower contrastive loss weight beneficial. In the shorter sessions of RetailRocket and Diginetica, higher contrastive loss weights help generate more comprehensive session representations by capturing intent from different views.


\begin{figure}[t]
 \centering
  \includegraphics[width=\columnwidth]{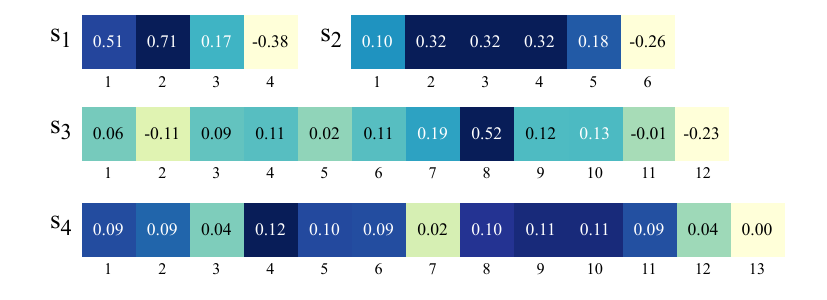}
  \caption{Multi-head attention visualization}
  \label{fig:attention_session}
\end{figure}

\begin{figure}[t]
 \setlength{\abovecaptionskip}{-2pt} 
 \centering
 \begin{subfigure}[b]{0.48\columnwidth}
  \includegraphics[width=\linewidth]{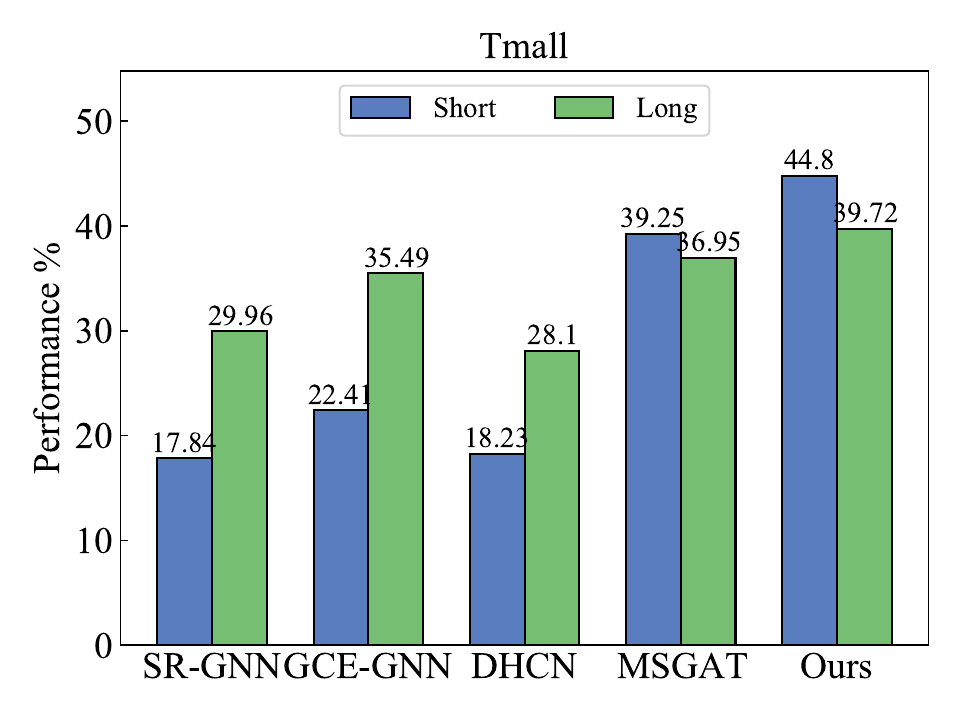}
  \end{subfigure}
  \hfill
  \begin{subfigure}[b]{0.48\columnwidth}
  \includegraphics[width=\linewidth]{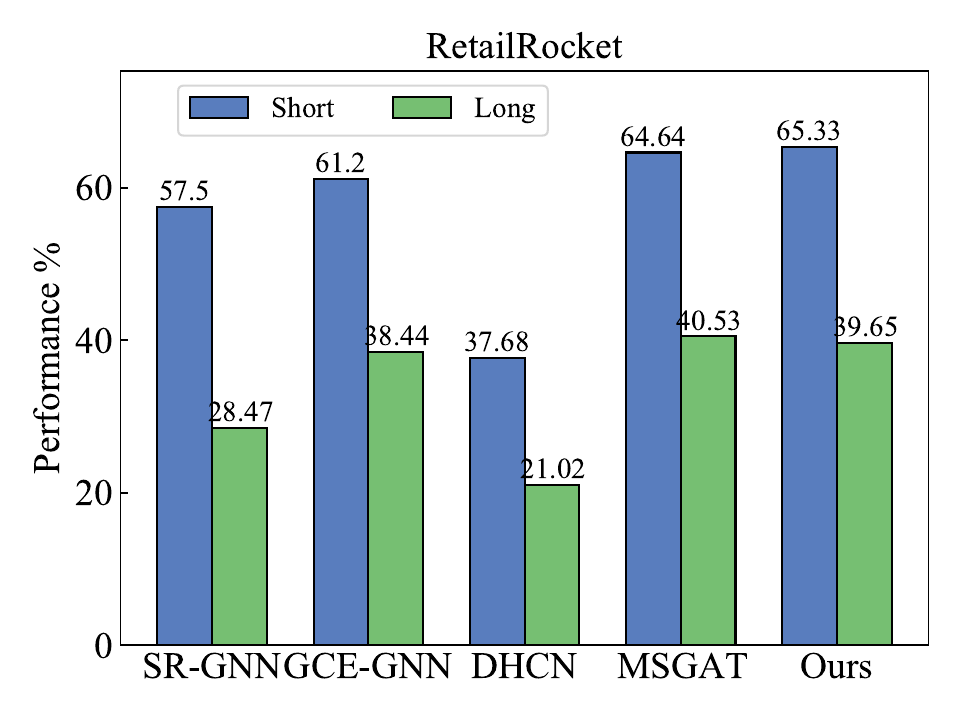}
  \end{subfigure}
  \caption{P@20 results on short and long sessions.}
  \label{fig:sessionLens}
\end{figure}

\subsection{Further Experiments}
In this section, we analyze the model's attention distribution and results for short and long sessions. \par
As shown in Figure \ref{fig:attention_session}, the importance of items is represented by the depth of color, with darker colors indicating higher importance. Based solely on the existing session sequence, it is difficult to directly determine the relationship between session items and the target item. However, through the visualization of the multi-head attention mechanism, we observe that the attention weight distribution varies across different sessions, reflecting the varying contributions of items in capturing user intent.
When the multi-head attention mechanism is removed, the evaluation metrics on the RetailRocket dataset show a significant decline, further highlighting the critical role of attention in capturing user intent. Therefore, beyond emphasizing the last item in the session, it is essential to dynamically learn and evaluate the influence of items at different positions on session intent. \par
We divide the Tmall and RetailRocket datasets into short and long sessions, with short sessions having 5 or fewer items and long sessions exceeding 5. We compare MGCOT with several representative baseline models, including SR-GNN, GCE-GNN, DHCN, and MSGAT. As shown in Figure \ref{fig:sessionLens}, MGCOT consistently outperforms these baselines across all session lengths, demonstrating its effectiveness in real-world session-based recommendation tasks.

\section{Conclusion}
This paper introduces the MGCOT model, which builds multiple graphs to capture the session intent from current, local, and global views. By integrating attention mechanisms, MGCOT effectively captures important information, while incorporating contrastive learning to generate more comprehensive and complementary session representations. Extensive experiments on three datasets demonstrate that our MGCOT model outperforms current SOTA models, validating its effectiveness in session-based recommendation tasks.

\section{Limitation}
The MGCOT model has several limitations. First, the construction of multiple graphs increases the storage space requirements. Second, the complexity of building self-supervised contrastive learning models leads to limited transferability and bulky model structures.

\section*{Acknowledgments}
This work was supported by the NSFC (62376180, 62176175, 62302329), the major project of natural science research in Universities of Jiangsu Province (21KJA520004), Suzhou Science and Technology Development Program (SYG202328, SKY2023128) and the Project Funded by the Priority Academic Program Development of Jiangsu Higher Education Institutions.
\bibliography{coling_latex}

\begin{thebibliography}{24}
\providecommand{\natexlab}[1]{#1}

\bibitem[{Chen et~al.(2023)Chen, Guo, Li, and Li}]{chen2023knowledge}
Qian Chen, Zhiqiang Guo, Jianjun Li, and Guohui Li. 2023.
\newblock Knowledge-enhanced multi-view graph neural networks for session-based recommendation.
\newblock In \emph{Proceedings of the 46th International ACM SIGIR Conference on Research and Development in Information Retrieval}, pages 352--361.

\bibitem[{Chen et~al.(2024)Chen, Wang, Fang, Meng, and Liang}]{chen2023heterogeneous}
Xiaoru Chen, Yingxu Wang, Jinyuan Fang, Zaiqiao Meng, and Shangsong Liang. 2024.
\newblock Heterogeneous graph contrastive learning with metapath-based augmentations.
\newblock \emph{IEEE Transactions on Emerging Topics in Computational Intelligence}, 8(1):1003--1014.

\bibitem[{Choi et~al.(2024)Choi, Kim, Cho, and Lee}]{choi2024multi}
Minjin Choi, Hye-young Kim, Hyunsouk Cho, and Jongwuk Lee. 2024.
\newblock Multi-intent-aware session-based recommendation.
\newblock In \emph{Proceedings of the 47th International ACM SIGIR Conference on Research and Development in Information Retrieval}, pages 2532--2536.

\bibitem[{Han et~al.(2022)Han, Zhang, Chen, Lai, Song, and Li}]{han2022multi}
Qilong Han, Chi Zhang, Rui Chen, Riwei Lai, Hongtao Song, and Li~Li. 2022.
\newblock Multi-faceted global item relation learning for session-based recommendation.
\newblock In \emph{Proceedings of the 45th international ACM SIGIR conference on research and development in information retrieval}, pages 1705--1715.

\bibitem[{Hidasi et~al.(2015)Hidasi, Karatzoglou, Baltrunas, and Tikk}]{hidasi2015session}
Bal{\'a}zs Hidasi, Alexandros Karatzoglou, Linas Baltrunas, and Domonkos Tikk. 2015.
\newblock Session-based recommendations with recurrent neural networks.
\newblock \emph{arXiv preprint arXiv:1511.06939}.

\bibitem[{Li et~al.(2017)Li, Ren, Chen, Ren, Lian, and Ma}]{li2017neural}
Jing Li, Pengjie Ren, Zhumin Chen, Zhaochun Ren, Tao Lian, and Jun Ma. 2017.
\newblock Neural attentive session-based recommendation.
\newblock In \emph{Proceedings of the 2017 ACM on Conference on Information and Knowledge Management}, pages 1419--1428.

\bibitem[{Liu et~al.(2018)Liu, Zeng, Mokhosi, and Zhang}]{liu2018stamp}
Qiao Liu, Yifu Zeng, Refuoe Mokhosi, and Haibin Zhang. 2018.
\newblock Stamp: short-term attention/memory priority model for session-based recommendation.
\newblock In \emph{Proceedings of the 24th ACM SIGKDD international conference on knowledge discovery \& data mining}, pages 1831--1839.

\bibitem[{Liu et~al.(2021)Liu, Zhang, Hou, Mian, Wang, Zhang, and Tang}]{liu2021self}
Xiao Liu, Fanjin Zhang, Zhenyu Hou, Li~Mian, Zhaoyu Wang, Jing Zhang, and Jie Tang. 2021.
\newblock Self-supervised learning: Generative or contrastive.
\newblock \emph{IEEE transactions on knowledge and data engineering}, 35(1):857--876.

\bibitem[{Ouyang et~al.(2023)Ouyang, Xu, Chen, Xie, Zheng, Song, and Zhao}]{ouyang2023mining}
Kai Ouyang, Xianghong Xu, Miaoxin Chen, Zuotong Xie, Hai-Tao Zheng, Shuangyong Song, and Yu~Zhao. 2023.
\newblock Mining interest trends and adaptively assigning sample weight for session-based recommendation.
\newblock In \emph{Proceedings of the 46th International ACM SIGIR Conference on Research and Development in Information Retrieval}, pages 2174--2178.

\bibitem[{Pan et~al.(2022)Pan, Cai, Chen, Chen, and Chen}]{pan2022collaborative}
Zhiqiang Pan, Fei Cai, Wanyu Chen, Chonghao Chen, and Honghui Chen. 2022.
\newblock Collaborative graph learning for session-based recommendation.
\newblock \emph{ACM Transactions on Information Systems (TOIS)}, 40(4):1--26.

\bibitem[{Qiao et~al.(2023)Qiao, Zhou, Wen, Zhang, and Gao}]{qiao2023bi}
Shutong Qiao, Wei Zhou, Junhao Wen, Hongyu Zhang, and Min Gao. 2023.
\newblock Bi-channel multiple sparse graph attention networks for session-based recommendation.
\newblock In \emph{Proceedings of the 32nd ACM International Conference on Information and Knowledge Management}, pages 2075--2084.

\bibitem[{Qiu et~al.(2019)Qiu, Li, Huang, and Yin}]{qiu2019rethinking}
Ruihong Qiu, Jingjing Li, Zi~Huang, and Hongzhi Yin. 2019.
\newblock Rethinking the item order in session-based recommendation with graph neural networks.
\newblock In \emph{Proceedings of the 28th ACM international conference on information and knowledge management}, pages 579--588.

\bibitem[{Ren et~al.(2019)Ren, Chen, Li, Ren, Ma, and de~Rijke}]{ren2019repeatnet}
Pengjie Ren, Zhumin Chen, Jing Li, Zhaochun Ren, Jun Ma, and Maarten de~Rijke. 2019.
\newblock Repeatnet: A repeat aware neural recommendation machine for session-based recommendation.
\newblock \emph{Proceedings of the AAAI Conference on Artificial Intelligence}, 33(1):4806--4813.

\bibitem[{Rendle et~al.(2010)Rendle, Freudenthaler, and Schmidt-Thieme}]{rendle2010factorizing}
Steffen Rendle, Christoph Freudenthaler, and Lars Schmidt-Thieme. 2010.
\newblock Factorizing personalized markov chains for next-basket recommendation.
\newblock In \emph{Proceedings of the 19th international conference on World wide web}, pages 811--820.

\bibitem[{Sun et~al.(2024)Sun, Guo, Zhang, Liu, and Wu}]{sun2024exploiting}
Tianqi Sun, Hongrui Guo, Zihan Zhang, Hongzhi Liu, and Zhonghai Wu. 2024.
\newblock Exploiting multifaceted nature of items and users for session-based recommendation.
\newblock In \emph{Proceedings of the 2024 SIAM International Conference on Data Mining (SDM)}, pages 580--588.

\bibitem[{Wang et~al.(2019)Wang, Ren, Mei, Chen, Ma, and De~Rijke}]{wang2019collaborative}
Meirui Wang, Pengjie Ren, Lei Mei, Zhumin Chen, Jun Ma, and Maarten De~Rijke. 2019.
\newblock A collaborative session-based recommendation approach with parallel memory modules.
\newblock In \emph{Proceedings of the 42nd international ACM SIGIR conference on research and development in information retrieval}, pages 345--354.

\bibitem[{Wang et~al.(2020)Wang, Wei, Cong, Li, Mao, and Qiu}]{wang2020global}
Ziyang Wang, Wei Wei, Gao Cong, Xiao-Li Li, Xian-Ling Mao, and Minghui Qiu. 2020.
\newblock Global context enhanced graph neural networks for session-based recommendation.
\newblock In \emph{Proceedings of the 43rd international ACM SIGIR conference on research and development in information retrieval}, pages 169--178.

\bibitem[{Wu et~al.(2019)Wu, Tang, Zhu, Wang, Xie, and Tan}]{wu2019session}
Shu Wu, Yuyuan Tang, Yanqiao Zhu, Liang Wang, Xing Xie, and Tieniu Tan. 2019.
\newblock Session-based recommendation with graph neural networks.
\newblock \emph{Proceedings of the AAAI Conference on Artificial Intelligence}, 33(1):346--353.

\bibitem[{Wu et~al.(2020)Wu, Pan, Chen, Long, Zhang, and Philip}]{wu2020comprehensive}
Zonghan Wu, Shirui Pan, Fengwen Chen, Guodong Long, Chengqi Zhang, and S~Yu Philip. 2020.
\newblock A comprehensive survey on graph neural networks.
\newblock \emph{IEEE transactions on neural networks and learning systems}, 32(1):4--24.

\bibitem[{Xia et~al.(2021{\natexlab{a}})Xia, Yin, Yu, Shao, and Cui}]{xia2021selfco}
Xin Xia, Hongzhi Yin, Junliang Yu, Yingxia Shao, and Lizhen Cui. 2021{\natexlab{a}}.
\newblock Self-supervised graph co-training for session-based recommendation.
\newblock In \emph{Proceedings of the 30th ACM international conference on information \& knowledge management}, pages 2180--2190.

\bibitem[{Xia et~al.(2021{\natexlab{b}})Xia, Yin, Yu, Wang, Cui, and Zhang}]{xia2021self}
Xin Xia, Hongzhi Yin, Junliang Yu, Qinyong Wang, Lizhen Cui, and Xiangliang Zhang. 2021{\natexlab{b}}.
\newblock Self-supervised hypergraph convolutional networks for session-based recommendation.
\newblock \emph{Proceedings of the AAAI Conference on Artificial Intelligence}, 35(5):4503--4511.

\bibitem[{Xu et~al.(2019)Xu, Zhao, Liu, Sheng, Xu, Zhuang, Fang, and Zhou}]{xu2019graph}
Chengfeng Xu, Pengpeng Zhao, Yanchi Liu, Victor~S. Sheng, Jiajie Xu, Fuzhen Zhuang, Junhua Fang, and Xiaofang Zhou. 2019.
\newblock Graph contextualized self-attention network for session-based recommendation.
\newblock In \emph{Proceedings of the Twenty-Eighth International Joint Conference on Artificial Intelligence, {IJCAI-19}}, pages 3940--3946.

\bibitem[{Zheng et~al.(2024)Zheng, Wu, Zhang, and Li}]{zheng2024hypergraph}
Xiangping Zheng, Bo~Wu, Alex~X Zhang, and Wei Li. 2024.
\newblock Hypergraph-based session modeling: A multi-collaborative self-supervised approach for enhanced recommender systems.
\newblock In \emph{Proceedings of the 2024 Joint International Conference on Computational Linguistics, Language Resources and Evaluation (LREC-COLING 2024)}, pages 8493--8504.

\bibitem[{Zhou et~al.(2020)Zhou, Wang, Zhao, Zhu, Wang, Zhang, Wang, and Wen}]{zhou2020s3}
Kun Zhou, Hui Wang, Wayne~Xin Zhao, Yutao Zhu, Sirui Wang, Fuzheng Zhang, Zhongyuan Wang, and Ji-Rong Wen. 2020.
\newblock S3-rec: Self-supervised learning for sequential recommendation with mutual information maximization.
\newblock In \emph{Proceedings of the 29th ACM international conference on information \& knowledge management}, pages 1893--1902.

\end{thebibliography}


\end{document}